\documentclass[onecolumn,showpacs,preprintnumbers]{revtex4}%
\usepackage{amsfonts}
\usepackage{amsmath,epsfig}
\usepackage{graphicx}
\usepackage{dcolumn}
\usepackage{bm}
\usepackage{amssymb}
\usepackage{amsmath}%
\begin{document}
\title{The merger of two compact stars: \\ a tool for dense matter nuclear physics}
\author{Alessandro Drago$^1$, Giuseppe Pagliara$^1$, Sergei B. Popov$^2$,
Silvia Traversi$^{1}$ and Grzegorz Wiktorowicz$^{3,4}$}
\affiliation{$^1$Dip.~di Fisica e Scienze della Terra dell'Universit\`a di Ferrara and INFN
Sez.~di
Ferrara, Via Saragat 1, I-44100 Ferrara, Italy}
\affiliation{$^2$ Sternberg Astronomical Institute, Lomonosov Moscow State University, Universitetsky prospekt 13, 119234, Moscow, Russia}

\affiliation{$^3$ National Astronomical Observatories, Chinese Academy of Sciences, Beijing 100012, China}
\affiliation{$^4$ School of Astronomy \& Space Science, University of the Chinese Academy of Sciences, Beijing 100012, China}
\begin{abstract}
 We discuss the different signals, in gravitational and electromagnetic waves, emitted during the merger of two compact stars. 
 We will focus in particular on the possible contraints that those signals can provide on the equation of state of dense matter. Indeed, the stiffness of the equation of state and the particle composition of the merging compact stars, strongly affect e.g. the life time of the post-merger remnant and its gravitational wave signal, the emission of the short gamma-ray-burst, the amount of ejected mass and the related kilonova. The first detection of gravitational waves from the merger of two compact stars in August 2017, GW170817, and the subsequent detections of its electromagnetic counterparts, GRB170817A and AT2017gfo, is the first example of the era of "multi-messenger astronomy": we discuss what we have learned from this detection on the equation of state of compact stars and we provide a tentative interpretation of this event, within the two families scenario, as due to the merger of a hadronic star with a quark star.
\end{abstract}

\maketitle

\section{Introduction}
The observation, on 2017 August 17, of the coalescence of two compact objects characterized by masses in the typical neutron star (NS) range has marked the beginning of the so called "multi-messenger astronomy" \cite{GBM:2017lvd}. Indeed, the merger event has provided a signal in gravitational waves (GW170817) detected by Advanced LIGO and Advanced VIRGO, that has allowed to localize the binary constraining a sky region of $31 \mathrm{deg}^2$ and a distance of a $40_{-8}^{+8}$ Mpc. Moreover, Fermi Gamma-ray Burst Monitor has detected a short Gamma- Ray-Burst event (GRB170817A) delayed by 1.7 s with respect to the merger time. These two detections have been followed by multiple observations revealing the existence of electromagnetic (EM) counterpart of the GW event covering the entire EM bands, with signals in the X, UV, optical, IR and radio parts of the spectrum.\\
The separated and joined analysis of these different signals can provide physical insides about open problems in theoretical physics and astrophysics which have been for years the subject of speculations and simulations.\\
In particular, the study of the optical counterpart of GW170817, called kilonova (AT2017gfo) because of its peak luminosity, has finally confirmed that NS mergers host r-processes responsible of the synthesis of the most heavy nuclei. Moreover, it has provided information about the amount and features of the ejecta and these could finally give constraints about the importance of the different ejection mechanisms and of the features of the progenitors.  

\section{State of the art before 17 August 2017}
\subsection{Expectations from the GW signal}
The merger of two compact stars represents one of the most powerful sources of GWs \cite{Shibata:2002jb}. The process of merger can be divided into three main stages: the inspiral phase, the coalescence phase and the post merger phase; each of these phases has its own specific waveform which in turn is determined by different physical quantities such as the total mass of the binary, the mass asymmetry, the spin of the two stars, the orbital parameters and, finally, the internal structure of the two stars.   
During the inspiral phase, the GW's signal is characterized by the chirp mass $ {\it M}=\frac{(m_1 m_2)^{3/5}}{(m_1+m_2)^{1/5}}$  where $m_1$ and $m_2$ are the masses of the two stars (it is customary to label with $m_2$ the smallest of the two masses). The detection of this part of the signal allows therefore to measure ${\it M}$ and to determine, with good accuracy, the total mass of the system $M=m_1+m_2$. This is due to the fact that from astrophysical observations and from supernovae numerical simulations one can infer that $m_2 \geq 1.1 M_{\odot}$. Similarly, the asymmetry parameter $q=m_2/m_1$ is likely to be larger than about 0.6.    
During most of the inspiral phase one can assume the two stars to be point like sources but when they are at a distance comparable to their radius their finite size can significantly modify the GW signal. Indeed, part of the potential energy of the binary 
is spent in perturbing the structure of the stars. In turn this leads to an
acceleration of the inspiral dynamics with respect to the case of point-like sources (or with respect to the case of a BH-BH merger). The physical quantity which parametrizes this effect is the tidal deformability $\Lambda$ of the two stars \cite{Flanagan:2007ix}. In general, at fixed mass, the larger the radius of the star the larger the value of $\Lambda$, the larger is the deviation of the GW signal from the case of point-like sources. Potentially, a precise measurement of the final part of the inspiral phase could lead to very interesting constraints on the radii of the merging compact stars.

Finally, let us discuss the outcome of the merger and 
the corresponding GW signal. A first possibility is that 
when the two compact objects merge a BH hole is formed promptly,
within a time scale of the order of 1ms. Correspondingly, the GW signal rapidly switches off. There have been many numerical studies on the conditions for obtaining a prompt collapse \cite{Baiotti:2008ra,Hotokezaka:2013iia,Bauswein:2013jpa,Bauswein:2017aur}. A remarkable result is that the value of $M$ above which the remnant collapses rapidly to a BH, $M_{\mathrm{threshold}}$, depends strongly on the equation of state of dense matter. 
In particular in Refs.\cite{Bauswein:2013jpa,Bauswein:2017aur}, it has been shown that the ratio between $M_{\mathrm{threshold}}$ and the maximum mass of the cold and non-rotating configuration $M_{\mathrm{TOV}}$ to good accuracy scales linearly with the compactness of the maximum mass configuration. 
This implies that once GW will be detected from mergers, this will allow to measure 
$M_{\mathrm{threshold}}$, and to obtain precious information on the structure of cold neutron stars and thus on the equation of state of dense nuclear matter.

If the remnant does not collapse immediately there are three possible outcomes of the merger:
a hypermassive star (i.e. a configuration which is stable only as long as differential rotation is not completely dissipated), a supramassive star (i.e. a configuration which is stable only as long as rigid rotation is present) and finally an initially differentially rotating star which is stable even without rotation. In all these cases, the remnant, during the so called phase of the ring-down, will also emit a powerful GW signal although with a spectrum qualitatively very different from the inspiral phase signal. In Refs.\cite{Bauswein:2011tp,Bauswein:2012ya,Takami:2014zpa,Maione:2017aux}, such spectrum has been 
studied as obtained from different numerical simulations and the dominant frequencies
have been singled out. Again, these frequencies depend strongly on the equation of state: potentially, if at least one of those modes could be detected one could constrain
the radius of the $1.6 M_{\odot}$ configuration within a few hundreds meters \cite{Bauswein:2015vxa}. One should notice however that in general the ring-down signal
lies in a frequency range above the kHz for which the sensitivity of LIGO and VIRGO is reduced.

\subsection{Mechanisms describing the prompt emission of short GRB and the
Extended Emission}
The problem of finding the inner engine of short GRBs is linked to the need of overcoming two difficulties: 
first, the generation of a jet with a large Lorentz factor implies a clean environment and therefore 
a mechanism able to reduce the baryonic pollution is needed; second, some but not all of the sGRBs display an
Extended Emission (EE), similar to the quasi-plateau emission observed in the case of long GRBs and lasting
up to $10^4$ s (or even more in a few cases), suggesting that
the inner engine does not switch-off completely after a fraction of a second. 

Concerning the way to reduce the
baryonic pollution, two mechanisms have been suggested: one is based on the formation of a Black-Hole, so that
baryonic material stops being ablated from the surface of the stellar object formed immediately after the merger \cite{Rezzolla:2011da};
the other suggested mechanism is based on the formation of a Quark Star (QS): also in this case baryonic material
cannot be ablated once the process of quark deconfinement has reached the surface of the star \cite{Drago:2015qwa}. 

Concerning 
the origin of the EE, again two mechanisms have been proposed. One is based on the formation of a proto-magnetar
and describes the EE in a way similar to the emission of a pulsar \cite{Lyons:2009ka,DallOsso:2010uxj,Rowlinson:2013ue}. 
One needs to assume that after the merger a supramassive star (or even a totally stable
star) is formed, since the collapse to a BH needs to be delayed at least by the time associated with the 
duration of the EE. 
This mechanism is able to reproduce in a very
accurate way the light-curves of the EE, just by using two parameters, the strength of the magnetic field 
(which needs to be of the order of about $10^{15}-10^{16}$ G, and the rotation period (which needs to be 
of the order of a few milliseconds, or shorter). The second mechanism is based on the formation of an
accretion disk around the BH \cite{vanPutten:2014kja}: although this possibility cannot be ruled out no attempt at modeling the
EE within this scheme as been made up to now. 

Since most of the sGRBs do not display any EE, it is quite natural to assume that most of them are associated
with the formation of a BH in less than a second and that in those cases no EE, due to an accretion disk, 
is produced. Assuming that the EE is explained via a protomagnetar model, two possibilities exist for
describing the sGRBs with EE. The first possibility assumes that the prompt emission is due
to the formation of a BH. Since the EE is observed after the prompt emission, this scenario
needs a "time-reversal" mechanism, so that the EE produced before the collapse to a BH is observed
after the prompt emission which is produced when the BH forms \cite{Rezzolla:2014nva,Ciolfi:2014yla}. The time-reversal is associated to
the time needed for the soft EE to leak out of the thick cocoon surrounding the protomagnetar. Instead the strong
prompt emission is emitted soon after the BH is formed and it exits the cocoon along the rotation axis.
The second possibility is that a QS forms, instead of a BH. In this way the prompt emission takes place when
the process of quark deconfinement has reached the surface of the star reducing the baryonic pollution and the
EE is due to the proto-magnetar that in this case is a QS \cite{Drago:2015qwa}. Notice that these two possibilities can be easily
distinguished by observations. The "time-reversal" mechanism implies that the prompt emission takes place
after the protomagnetar collapses to a BH and therefore the time-separation between the moment of the merger (observed
in GWs) and the prompt emission (observed in x- and $\gamma$-rays) is of the order of the duration of a supramassive
star, i.e. it is easily larger than $10^3 - 10^4$ s. Instead the mechanism based on the formation of a QS
requests a time separation between merger and prompt emission of the order of about 10s, needed for the deconfinement
front to reach the surface of the star. This a relevant example 
of multimessenger analysis at the base of proposals such as 
the THESEUS mission \cite{Stratta:2017bwq}.

\subsection{Ejected mass from NS mergers, r-processes and EM signal}

The question about the correct astrophysical mechanism that could be at the base of the r-process nucleosynthesis represents one of the subject on which the physicists are focusing on in the last decade.
The first attempt to explain the mystery was to indicate the process of core collapse supernovae (CCSN) as the ideal environment in which r-processes could take place \cite{Burbidge:1957vc}.
But, recently, detailed calculations,  have shown that CCSN don't appear to host the right conditions to create the most neutron-rich nuclei \cite{Hoffman:2007du, Fischer:2009af,Arcones:2006uq,Roberts:2010wh}. In particular, it seems to be especially difficult for core-collapse supernovae to produce what is known as the "third-peak". 
These results have pushed the researchers to try to find other possible astrophysical sources which can be responsible for a sufficient emission of matter in the right conditions for r-processes to happen. 
In the following we discuss the possibility that r-processes take place during the merger.

\subsubsection{Ejection mechanism and features of the outgoing fluid} 

Binary neutron star (BNS) mergers can result in the ejection of neutron-rich matter, by means of several different possible processes. A classification of the different components of the ejecta has already been made in 2015 by Hotoketzaka and Piran \cite{Hotokezaka:2015eja}: the main sources are a dynamical ejection and a later ejection of part of the disk formed around the remnant because of neutrino or viscous heating. 

The dynamical ejection is due to two different physical mechanisms: the first one is the tidal deformation of the NS, a consequence of the gravitational field that is not axisymmetric; the matter gains sufficient angular momentum and the ejection, mostly in the equatorial plane, starts before the collision and ends about $10\;ms$ after the merger \cite{Hotokezaka:2012ze}. This material is characterized by a very low electron fraction, $Y_e<0.1$ \cite{Palenzuela:2015dqa,Radice:2016dwd} which can eventually be increased by means of weak reactions in few ms after the merger \cite{Sekiguchi:2015dma}. 
The second is the shock that is formed at NSs interface, that spreads the crust material. Also, in the envelope of the remnant, a shock is produced by radial oscillations giving to some fraction of matter the sufficient energy to be ejected. The shock component could be dominant in the case of equal-mass binaries and can be ejected also in the polar direction. The electron fraction is predicted to be higher with respect to the tidal component with values in the range $0.2<Y_{e}<0.4$ \cite{Palenzuela:2015dqa,Radice:2016dwd,Sekiguchi:2015dma}. This difference is caused principally by the higher neutrino flux which characterizes the polar direction with respect to the equatorial one: indeed electron (anti)neutrino, electron and positron captures can have a deep influence on the evolution of the electron fraction of the ejecta \cite{Goriely:2015fqa}.
The dynamically ejected fluid is characterized by a velocity that can reach values of $\beta\sim 0.2-0.3$.  

After the merger, some of the ashes of the NSs surround the central part forming a disk of mass in the range $10^{-3} M_{\odot} <M_{disk} < 0.3 M_{\odot}$ \cite{Shibata:2006nm,Rezzolla:2010fd,Hotokezaka:2013iia}. Part of this disk can generate an outflow caused by viscous or neutrino heating, whose features are characteristic of the type of remnant.  If the remnant is a NS the outflow depends also on its lifetime. The strong magnetic fields present at this stage can also play a role. The amount of ejecta is estimated to vary from 5 to 20$\%$ of the mass of the disk. This ejecta is usually characterized by lower velocities with respect to the dynamical one  reaching a maximum value of about 0.1c \cite{Siegel:2017nub}. The electron fraction of this type of ejecta, initially quite low ($\sim 0.1$), can be significantly modify by neutrinos, finally spanning in a range $0.05 -0.5$ with a distribution which depends on the equation of state employed \cite{Palenzuela:2015dqa}. 

Many general relativistic (GR) hydrodynamical simulations of the merger have been performed in order to describe the features of these different kind of ejecta, to evaluate the total amount of material expelled during the phases of the merger and to study the dependence of the results on the features of the binary and on the equation of state describing the NSs.

Concerning the dynamically ejected mass,
its tidal component depends on the tidal deformability 
$\Lambda$: the stiffer the equation of state, the larger the value of $\Lambda$ and the larger the amount of tidally ejected mass. 
On the other hand the compactness of the stars can influence the shock produced at the merger, and also the quantity of material that can be spread out at the moment of the merger. Soft EOSs determine a larger impact velocity and so it is plausible that the correspondent shock and ejected mass will be higher \cite{Bauswein:2013yna}.
Concerning the amount of mass ejected from the disk, it is limited by the mass of the disk which in turns depend on the life time of the hypermassive star.
Therefore, this component is larger for stiffer equations of state \cite{Perego:2017wtu}.
Future detections of kilonovae will allow to disentagle the various 
components providing crucial information for nuclear physics \cite{Stratta:2017bwq}.

Finally, the amount of ejected matter is deeply influenced by the degree of asymmetry of the binary $q$. For more asymmetric binaries, the unbound material is larger than in the symmetric case. This result can be explained in terms of the bigger effect of the tidal force that cause the lighter star to be deformed to a drop-like object and, after the merger, to be stretched leading to the formation of a pronounced tidal tail.
Also the average electron fraction is influenced by the degree of asymmetry: the effect is particularly strong for soft EOS and it manifests itself as a decrease of the electron fraction of the ejecta with the increase of the mass asymmetry \cite{Sekiguchi:2016bjd}.
\subsubsection{R-processes}
The ejected mass is reprocessed and through r-processes can in principle 
generate the distribution of heavy nuclei. It is an open question whether 
NSs mergers eject an amount of matter sufficient to explain the observed abundances. For these reasons, plenty of simulations have been performed in order to reproduce the path of r process nucleosyntesis: the reaction network included nuclear species between the stability valley and the neutron drip line and considered neutron captures, photodisintegration reactions together with fission and $\beta-$decay reactions \cite{Goriely:2011vg,Korobkin:2012uy,Goriely:2013eua,Bauswein:2013yna,Just:2014fka,Siegel:2017jug}.  
The comparison between the solar abundances as a function of mass number $A$ and the results of these simulated nucleosyntesis (in which the quantity of ejected mass is of the order of $10^{-3}-10^{-2}\; M_{\odot}$ and the merger rate for galaxy is set in a range $10^{-5} - 10^{-4}\; yr^{-1}$) shows a good agreement in the regime $A>120 - 140$, i.e. a region corresponding to the second and the third peak.

The ability of the simulations to reproduce the abundances of the elements lying between the first and the second peak depends on the obtained distribution of the electron fraction of the ejecta and on the inclusion of the entire network of possible weak interactions. Indeed in Refs. \cite{Goriely:2011vg,Korobkin:2012uy, Bauswein:2013yna} only the dynamically ejecta are considered while in \cite{Siegel:2017jug} the outflow of material from the disk is also studied, but all the simulations fail in reproducing the abundances for $A<120$ because of the low electron fraction attributed to ejecta as a result of neglecting the neutrino absorption processes.
In \cite{Goriely:2015fqa} the authors include also the weak interaction of free neutrons obtaining a significant fraction of material with $Y_e=0.3-0.4$ responsible for the production of nuclei in the range $A = 90 -140$.

\subsubsection{EM counterpart}
A probe of the amount of ejected mass and of the realization of the r-process chains in NS mergers can be the analysis of the EM signal predicted to be associate with this phenomenon \cite{Metzger:2010sy}. 
The maximum of the luminosity takes place just after the photons can escape the expanding ejecta whose density is reducing.
A typical timescale is of the order of 1 day while the luminosity $\sim 10^{42}$ erg s$^{-1}$, three order of magnitude  larger than the Eddington luminosity for a solar mass star: for this reason this EM events are called kilonovae.
The spectral peak can vary in the IR/optical/nearUV wavelenghts.
The timescale $t_{peak}$, the luminosity $L_{peak}$ and the effective temperature $T_{peak}$ of the signal depend on the amount $M_{ej}$, the velocity  $v$ and the opacity $k$ of the ejecta \cite{Metzger:2010sy}:
\begin{equation}
t_{peak}\propto\;\biggl( {\frac{k M_{ej}}{v}}\biggr)^{\frac{1}{2}} \;,\;\; L_{peak}\propto\;\biggl( {\frac{v M_{ej}}{k}}\biggr)^{\frac{1}{2}} \;,\;\;  T_{peak}\propto\;\biggl( {vM_{ej}}\biggr)^{-\frac{1}{8}}k^{-\frac{3}{8}}\nonumber
\end{equation}
These dependences on the features of the ejecta could translate in an influence of the EOS of NS: an EOS which produce more ejecta will lead to a brighter optical counterparts, peaked on longer timescales and with longer peak wavelengths \cite{Bauswein:2013yna}.

\section{GW170817-GRB170817A-AT2017gfo}
\subsection{Analysis of the GW signal}
The signal detected by the LIGO-VIRGO collaborations \cite{TheLIGOScientific:2017qsa} corresponds to the emission of GWs from an inspiral binary with a chirp mass ${\it M}=1.188^{+0.004}_{-0.002} M_{\odot}$ which implies a total mass $M=2.74^{+0.04}_{-0.02}M_{\odot}$ (under the hypothesis that the spins of the two stars are compatible with the ones observed in binary neutron stars, "low spin case"). In turn, the masses of the components are in the range $1.17-1.6 M_{\odot}$, strongly suggesting that the merger was between two NSs. Although the source is quite close, $40$ Mpc, it has not been possible to follow the GW signal up to the merger and during the ring-down phase. However, a very useful upper limit on the value of the tidal parameter $\tilde{\Lambda}$ (which depends on the tidal deformabilities and the masses of the two stars) has been set: $\tilde{\Lambda}<800$ at the $90\%$ level in the low spin case. This constraint is basically model independent and it allows to already rule out a few very stiff equations of state such as MS1 and MS1b which are based on relativistic mean field calculations \cite{Mueller:1996pm}. 

What happened during the first few milliseconds after the merger is unclear.
Even if not completely excluded, the possibility that the merger has led to a prompt collapse seems to be very unlikely beacause in that case it would be difficult to explain the observation of the electromagnetic counterparts of GW170817. Actually, one can infer that the post-merger remnant is most probably a hypermassive star: a supramassive star or a stable star would inject part of its huge kinetic rotational energy into the GRB or into the kilonova on a long time scale and there is no evidence, in the observed signals, of such an energy injection \cite{Margalit:2017dij}.
This implies that the total energy of the binary, which can be estimated to be of the order of $95\%M$ (assuming the gravitational binding energy of the binary to be $\sim 5\%M$) is larger than 
the maximum mass of the supramassive configuration $M_{\mathrm{supra}}$. Several numerical calculations on rotating compact stars have shown that $M_{\mathrm{supra}}$ is to good accuracy $\sim 1.2 M_{\mathrm{TOV}}$ \cite{Lasota:1995eu}. Combining these results one therefore obtains that $M_{\mathrm{TOV}} < 2.2 M_{\odot}$.
This simple estimate is in agreement with the results of Refs.\cite{Margalit:2017dij,Ruiz:2017due,Rezzolla:2017aly}
and it again disfavors very stiff equations of state which predict 
maximum masses above $2.2 M_{\odot}$ such as e.g. DD2 \cite{Banik:2014qja}.

If the remnant is a hypermassive star, another constraint can be obtained 
by imposing that the total energy of the binary is lower than the maximum mass of the hypermassive configuration. This study has been performed in \cite{Bauswein:2017vtn} and it allows to rule out extremely soft equations of state: it has been found that the radius of the $1.6M_{\odot}$ configuration must be larger than about $10.7$ km.

To summarize, the first detection of GWs from binary 
neutron stars has already allowed to exclude a few examples
of dense matter equations of state. In particular, very stiff 
equations of state based only on nucleonic degrees of freedom 
seem to be unfavored. We will discuss in the last section
how this result actually suggests that strange matter
must appear in some form in compact stars.
\subsection{The weak gamma emission of GRB170817A: was it a standard short GRB?}
As already discussed above, short GRBs are assumed to originate from the merger of two NSs. In the case of the event of August 2017 the GW signal clearly indicates that a merger did take place but, on the other hand, the gamma-ray emission was delayed by approximately two seconds respect to the moment of the merger and the observed signal was much weaker than the one of a typical short GRB. It is also relevant to stress that no extended emission was observed, likely indicating that a supramassive star did not form after the merger.

There are two main possible interpretations of the event. The first one assumes that the emission was intrinsically sub-luminous and quasi-isotropic 
\cite{Gottlieb:2017pju,Kasliwal:2017ngb}. The second one assumes instead a standard short GRB emission, that was observed off-axis \cite{Lazzati:2017zsj}. While at the moment, about a hundred days after the event, both possibilities can explain the data, the analysis of the future time-evolution of the emission will ultimately be able to distinguish between these two scenarios, telling therefore if GRB170817A was a standard short GRB seen off-axis or if it belongs to a new class of phenomena \cite{Margutti:2018xqd}.

Even though at the moment the mechanism which launched GRB190817A is still unclear, some strongly energetic emission in $\gamma$ and in x-rays was produced and this indicates that the merger did not collapse instantaneously to a BH. There are explicit simulations indicating that if a jet needs to be formed the object produced in the post-merger needs to survive for at least a few tens of milliseconds \cite{Ruiz:2017inq}. As discussed in the following, also the analysis of the kilonova emission indicates that the result of the merger did not collapse immediately to a BH: a relevant amount of matter was likely emitted from the disk on a time-scale incompatible with an almost instantaneous collapse. This is a very important point to take into account when discussing the possible models for the merger, as we will do in the last section.
\subsection{Electromagnetic signal and mass ejection}
\subsubsection{Analysis of the optical transient}
On 2017 August 17 there has been the observation of the first electromagnetic counterpart to a gravitational wave event attributed to a merger of two NSs.
The data in UV, optical and NIR bands extend for a time interval from 0.47 to 18.5 days after the merger and are consistent with a kilonova signal predicted to be associated with a NS merger.

The early spectrum is dominated by a blue component. Over the first few days the spectra shows a rapid evolution to redder wavelengths: at 1.5 days after merger the optical peak is located around 5000 A and already at 2.5 days it shows a shift to 7000 A, evolving to $\sim$ 7800 A at 4.5 days and going finally out of the optical regime in the interval between 4.5 and 7.5 days after the merger. By 10 days the wavelength is $>$ 15000 A.  \cite{Nicholl:2017ahq}.  Moreover, the rate of the decline is observed to change for the different bands belonging to the observed kilonova spectrum: while the decline appears to be quick in the ug band (blue) with a rate of $\sim$ 2 mag day, the rizY (red) and the HKs (NIR) bands show a smoother decay causing the spectrum to be dominated by red at late time \cite{Cowperthwaite:2017dyu}. 

The initial luminosities, $\sim 5\cdot 10^{41}$ erg s$^{-1}$ at 0.6 days and $\sim 2\cdot 10^{41}$ erg s$^{-1}$ at 1.5 days, and the short timescale ($\sim$ 1 day) are consistent with the model called Blue KN: this kind of emission was first proposed by Metzger and al. \cite{Metzger:2010sy} and subsequently developed as the signal associated to different kinds of matter likely to be ejected during or post merger \cite{Roberts:2011xz,Metzger:2014ila}: in \cite{Roberts:2011xz} the authors analyzed the tidal tails formed during the merger while in \cite{Metzger:2014ila} is presented a study of the outflow from the remnant in the case of a delayed ($>$100ms) BH formation. All this analysis have in common the very low opacity attributed to the ejected mass with values in a range from $k = 0.1$ to $k=1$ cm$^{2}$s$^{-1}$, typical of material containing Fe-group or light r-process nuclei characterized by $A<140$. Therefore, the Blue KN signal  is likely to be associated with r-processes responsible of the formation of nuclei lying between the first and the second peak.

Conversely, the late EM emission, which dominates at longer timescales $\sim$ a week and shows a lower luminosity $\sim$ $10^{40}-10^{41}$erg s$^{-1}$ fits well with the so called Red KN model \cite{Cowperthwaite:2017dyu}: in \cite{Barnes:2013wka}, \cite{Kasen:2013xka}, \cite{Tanaka:2013ana} and \cite{Tanaka:2017lxb} it was first presented the study of the effect of higher opacity of the ejecta on the resulting KN emission. This high opacity (k up to 10 cm$^{2}$s$^{-1}$) is attributed to the presence of Lanthanide elements, heavy nuclei with $A>140$, so the Red KN represents an indication of nucleosyntesis reactions filling the third peak of r-processes.

These observational evidences suggest the presence of material characterized by a not unique value of the opacity and therefore a different content of Lanthanides.
Despite a single component ejecta with a power-law velocity distribution and a time-dependent opacity (studied with an analytical model in \cite{Waxman:2017sqv}) can not be excluded, the most accredited hypothesis is the existence of at least two component of the ejecta, a Lanthanide poor (for the Blue KN) and a Lanthanide rich (for the Red KN) component \cite{Metzger:2014ila,Wollaeger:2017ahm}. This conclusion is also suggested by the fact that the blue component is not obscured by the red one, a clue of the need of distinct regions and angles of emission for the material with different opacity values. This also means that these two components can be attributed to distinct sources  \cite{Cowperthwaite:2017dyu}.

The duration and effective temperature of the KN emission have been studied with models outlined in \cite{Kasen:2013xka} and \cite{Villar:2017oya} allowing to indicate the mass, velocity and opacity of the ejecta as fitting parameters. The result is that the signal is consistent with a two component model. It consists of: a Blue component with $M^B_{ej}\sim 0.01 -$few $ 0.01\; M_{\odot}$, velocity $v^B_{ej}=0.27-0.3$c and opacity $k^B= 0.5$  cm$^{2}$s$^{-1}$ requiring a Lanthanide fraction of $\sim 10^{-4}$ to $10^{-5}$ in the outermost ejecta \cite{Cowperthwaite:2017dyu,Nicholl:2017ahq};  a Red component  with $M^R_{ej}\sim$ 0.04 $ M_{\odot}$, velocity $v^R_{ej}=0.12$c and opacity $k^R= 3.3$  cm$^{2}$s$^{-1}$ requiring a Lanthanide fraction of $\sim 10^{-2}$  \cite{Cowperthwaite:2017dyu,Chornock:2017sdf}.

Another model able to fit the data is characterized by three components: a Blue one with $M^B_{ej}\sim$ 0.01 $ M_{\odot}$, velocity $v^B_{ej} = 0.27$c and opacity $k^B= 0.5$  cm$^{2}$s$^{-1}$; a Purple one with $M^P_{ej}\sim$ 0.03 $ M_{\odot}$, velocity $v^P_{ej}=0.11$c and opacity $k^P= 3$  cm$^{2}$s$^{-1}$ and a Red one with $M^R_{ej}\sim$ 0.01 $ M_{\odot}$, velocity $v^R_{ej}=0.16$c and opacity $k^R= 10$  cm$^{2}$s$^{-1}$  \cite{Cowperthwaite:2017dyu}.

\subsubsection{Role of different ejection mechanisms}

The different opacities of the Red and Blue (and eventually Purple) KN, attributed to a different Lanthanide fraction can be directly connected to the electron fraction of the ejected matter, $Y_e$: a lower electron fraction corresponds to the ability to synthesize heavier nuclei, so to a bigger concentration of Lanthanides and, as a consequence to a greater opacity. The $Y_e$ of the ejecta depends
in turn on the direction and of the mechanism at the base of the ejection \cite{Wollaeger:2017ahm}.

The Blue component KN, characterized by a very low opacity, has to find its origin in a Lanthanide-poor material with an electron fraction $>0.25-0.3$: this kind of matter can be ejected dynamically by means of the shock generated at the contact surfaces of the two stars at the moment of the merger \cite{Palenzuela:2015dqa,Sekiguchi:2015dma,Goriely:2015fqa,Bauswein:2013yna}; the ejecta is expected to be be found within an angle between 30$^{\circ}$ and 45$^{\circ}$ \cite{Sekiguchi:2016bjd,Radice:2016dwd}, with respect to the polar axis where the neutrino flux is more intense and so neutrino absorption play a central role in raising the electron fraction above a value of $Y_e>0.25$ \cite{Kasen:2013xka,Perego:2017wtu}.
Moreover, if the remnant of the merger survives as an HMNS, a disk is formed around it which reaches a stable configuration in a few tens of milliseconds while a neutrino-driven wind is formed at a time of $\sim$ 10 ms \cite{Perego:2014fma}. This wind is responsible for the ejection of about the 5$\%$ of the mass of the disk, mostly in the polar direction. Simulations reveals that in this direction the large neutrino flux raises the electron fraction up to a distribution which peaks at $Y_e = 0.3 -0.4$, so wind also gives rise to a low-opacity ejecta \cite{Perego:2014fma,Fernandez:2013tya}.
The features which mostly distinguish the two different mechanism is the resulting velocities, 0.2-0.3 c for the dynamical ejecta while lower for the wind ejecta, v$< 0.1 c$ \cite{Siegel:2017nub,Perego:2014fma,Fernandez:2013tya}.

The velocity attributed to the Blue component, 0.27 - 0.3 c, represents an important proof of its dynamical origin, but while some authors suggest that the shock represents the exclusive mechanism for this low opacity signal \cite{Cowperthwaite:2017dyu,Nicholl:2017ahq}, others view this component as a possible result of a union of the dynamical and wind ejecta \cite{Perego:2017wtu}.
In the first case, the required amount of ejected matter, $\sim$ 10$^{-2}\;M_{\odot}$ implies the need for a soft EOS in order to reach an high velocity at the impact of the two compact objects: this suggest an upper limit on the NS radius of about $12$ km or less \cite{Nicholl:2017ahq}.
The second hypothesis, instead, does not require such a tight limit on the radius.

For what concerns the higher opacity, Lanthanide richer Red component the largely accepted interpretation indicates the dynamical mechanism of tidal ejection in the equatorial plane (within an angle of 45$^{\circ}$ - 60$^{\circ}$). The squeezed out material is indeed characterized by a very low electron fraction $<0.1$  \cite{Hotokezaka:2012ze,Palenzuela:2015dqa,Radice:2016dwd} giving rise to a Red-NIR spectrum with a longer timescale \cite{Kasen:2013xka,Barnes:2013wka}. The large amount of mass inferred from the data can be an indication of an high degree of asymmetry of the binary \cite{Cowperthwaite:2017dyu}.
However, to explain the component characterized by an opacity of $\sim$ 3 cm$^2$g$^{-1}$ and a very large ejected mass, which can be considered as part of the Red KN or a distinct Purple KN, it is necessary to take into account also the disk outflow.

First of all the wind ejecta for angles $>$30$^{\circ}$ are less affected by the neutrino flux maintaining an electron fraction $Y_e\sim$0.25-0.3 and fitting the required opacity \cite{Tanaka:2017lxb,Perego:2014fma,Kasen:2014toa}. At the same time, a contribution can also come from the secular ejecta which affects all the solid angle, but which is equatorial dominated: this viscous-driven ejection can results in the expulsion of up to 30$\%$ of the mass of the disk and the $Y_e$ of the material depends on the lifetime of the HMNS with respect to that of the disk ($\sim$ ten of ms). In the case of a long-lived HMNS the electron fraction can reach values between 0.2 - 0.5 with peaks at $\sim$ 0.3 - 0.4 while if the collapse to black hole happen earlier $Y_e<$0.3 -0.4 \cite{Fujibayashi:2017puw}.

 In \cite{Perego:2017wtu} ( three component model ) and in \cite{Cowperthwaite:2017dyu} the authors suggest that the intermediate opacity component of the KN signal can be, indeed, explained by means of the early viscosity driven secular ejection: this will imply a short-lived remnant ($\sim$ 30 ms) and a massive disk $\sim$ 0.08 $M_{\odot}$. 
 These two statements point to different directions concerning the features of the EoS: on one side a soft EOS will prevent the remnant to form a long-lived massive neutron star, but on the other side, the greater value of tidal deformability associated to a stiffer EOS will determine the formation of more pronounced tidal tales and thus a more massive disk around the remnant. On the other hand, the upper limit imposed on the tidal deformability by the gravitational waves measurement (see section 3.1 for details) and the absence of a prompt collapse to BH exclude extremely stiff or extremely soft EoS, respectively.
This seem to suggest an EoS characterized by an intermediate softness.

To summarize, the situation concerning the mechanisms at the base of the kilonova (and of the GRB) is still not settled. In the following we will shortly discuss the global interpretation of the event of August 2017 at the light of the two-families scenario.

\section{A different hypothesis: a hadronic star - quark star merger}
The event GW170817 and its electromagnetic counterparts have been generated from the coalescence of two compact stars. In the standard scenario, only one family of compact stars does exist, namely the family of stars composed entirely by hadronic degrees of freedom. However, there are some phenomenological indications of the possible existence of a second family of compact stars which are entirely composed by deconfined quarks, namely QSs, see Refs. \cite{Drago:2015cea,Drago:2015dea,Wiktorowicz:2017swq}.
In this scenario, the first family is populated by hadronic stars (HSs) which could be very compact and "light" due to the softness of the hadronic EoS (with hyperons and delta resonances included) while the second family is populated by QSs which, on the other hand, can support large masses due to the stiffness of the quark matter EoS.

In this scheme, a binary system could be composed by two HSs, by two QSs or finally by a HS and a QS.
Let us discuss these three possibilities in connection 
with the phenomenology of GW170817.

The threshold mass $M_{\mathrm{threshold}}$ for a HS - HS, i.e. the limit mass above which a prompt collapse is obtained, has been estimated to be $\sim 2.7 M_{\odot}$ \cite{Drago:2017bnf}, on the base of the the study performed in \cite{Bauswein:2015vxa}. This value is smaller than the total binary mass $M$ inferred from GW170817 \cite{TheLIGOScientific:2017qsa} and therefore the hypothesis that the binary sytems was a HS-HS system
is disfavored within the two families scenario.
Also, the possibility that the system was a double QS binary system is excluded because in that case 
it would be difficult to explain the kilonova which is powered by nuclear radioactive decays: even if some material is ejected from the QSs it is not made of ordinary nuclei and therefore cannot be used inside a r-process chain to produce heavy nuclei.
Conversely, the case of a HS - QS merger, in which the prompt collapse is avoided by the formation of a hypermassive hybrid configuration, becomes the most plausible suggestion in the context of the two families scenario  \cite{Drago:2017bnf}. 

Let us briefly discuss which are the possible evolutionary paths that can lead to the formation of such a mixed 
system.

The formation of a double compact object like the source of GW170817 is more probable in isolated binary evolution than through dynamical interactions in dense stellar systems \citep{Belczynski1712}. In such a case, a common envelope phase \citep{Ivanova1302} is typically necessary to shorten the orbital period and to allow for the merger to occur within the Hubble time. This phase, in a typical double NS formation route, occurs after the formation of the first compact object  \citep{Chruslinska1803}. Additionally, the companion may fill its Roche lobe again due to an expansion on the Helium main-sequence and commences a mass transfer phase. Therefore, it is natural to expect that the compact object which formed first may obtain a significantly higher mass then its counterpart. If the total mass of a double compact object is about $2.7M_{\odot}$ or higher, the heavier compact object may reach the mass of $\sim1.5\div1.6M_{\odot}$, which in the two-families scenario marks the threshold for deconfinement and the formation of a QS \citep{Drago:2015cea}. The secondary, cannot accrete mass from the primary, which is already a compact star. Moreover, it has lost a large fraction of mass in the interaction. Therefore, its pre-SN mass will be relatively low and, consequently, its post-SN mass will not be significantly different from the lower limit of a newborn NS $\sim1.1 M_{\odot}$; see e.g. \cite{Fryer1204}. This implies that in the two-families scenario, the binary evolution may favor the situation in which GW170817 is a HS-QS binary rather than a HS-HS. This issue will be further investigated in a forthcoming paper \cite{new-paper}.

Under the hypothesis that the event seen in August 2017 is due to the merger of a HS-QS system,
we need now to discuss the possible explanations of the different features seen in the gravitational and electromagnetic signals.
First, the gravitational wave signal has clearly indicated that extremely stiff EoSs are ruled out: the limit put on $\tilde{\Lambda}$ is fulfilled only if the radii of the two stars are smaller than about $13.4$km (see the analysis of Ref.\cite{Annala:2017llu}).
Both HSs and QSs satisfy this limit \cite{Drago:2013fsa,Drago:2015cea,new-paper}, see also Ref.\cite{Hinderer:2009ca} where the tidal deformabilities of HSs and QSs have been computed. 

Second, the emission of GRB170817A is probably connected with the formation of a relativistic jet which is launched by a BH-accretion torus system. The scenario discussed in \cite{Drago:2015qwa} concerns short GRB featuring and extended emission which has not been observed in this case. In our scenario, the compact star which forms immediately after 
the merger is a hypermassive hybrid star in which the burning of hadronic matter is still active. We expect such a system to collapse to a BH once the differential rotation is dissipated. The sGRB would be produced by the same mechanism studied in Refs.\cite{Rezzolla:2011da,Ruiz:2016rai}.

Let us finally discuss the properties of the observed kilonova within our scenario. Perego et al. \cite{Perego:2017wtu} suggest  an effective two components model in which the opacity of the secular ejecta is predicted to be very low ($\sim$ 1  cm$^{2}$s$^{-1}$), comparable to that of the wind component. This hypothesis has two major consequences: the lifetime of the remnant must be sufficiently long in order to allow weak reactions to raise the electron fraction to $> 0.3$ and the tidal ejecta must give a very relevant contribution.

Both these requirements can be fulfilled in the context of the HS - QS merger; indeed the hybrid star configuration predicted by this model can survive as a hypermassive configuration for a time of the order of hundreds of ms. Moreover, for an asymmetric binary, characterized by $q=0.75 - 0.8$, the predicted tidal deformability of the lightest star (the hadronic one) can reach value of $\sim 500$ \cite{new-paper}. This quite high value of $\lambda$ together with the supposed high asymmetry of the binary can result in a relevant contribution of the tidal effect on the total ejected mass. This allows to explain the third peak of r-processes and the Red KN without the need of a high opacity secular ejecta (notice also that the value $\lambda \sim$ 500 is largely above the lower limit derived from the analysis of the EM counterpart performed in \cite{Radice:2017lry}).

In conclusion, despite the need of hydrodynamical simulations in order to make more quantitative predictions, the HS-QS merger can represent a viable way to explain the features of the GW170817, GRB170817A and AT2017gfo.

\section*{Acknowledgments}
S.P. acknowledges support from the Russian Science Fondation project 14-12-00146. G.W. is partly supported by the President's International Fellowship Initiative (PIFI) of the Chinese Academy of Sciences under grant
no.2018PM0017 and by the Strategic Priority Research Program of the
Chinese Academy of Science “Multi-waveband Gravitational Wave Universe”
(Grant No. XDB23040000).


\end{document}